# Development of a Digital Twin for an Electric Vehicle Emulator Modeling, Control, and Experimental Validation

Lamine Chalal[1] and Ahmed Rachid[2]

*Abstract*— This paper presents the development and validation of a digital twin for a scaled-down electric vehicle (EV) emulator, designed to replicate longitudinal vehicle dynamics under diverse operating conditions. The emulator integrates a separately excited DC motor (SEDCM), a four-quadrant DC-DC converter, a battery emulator, and a mechanical load emulator. The system models tractive effort, aerodynamic drag, and gradient resistance using Newton's second law. In contrast to conventional graphical modeling tools (e.g., block diagrams and bond graphs), the adopted Energetic Macroscopic Representation (EMR) framework offers clear advantages by explicitly representing energy interactions and facilitating the systematic derivation of control structures. A control strategy developed within this framework governs energy flow across the powertrain, enabling accurate speed control via armature voltage regulation. Experimental tests conducted on a Lucas-Nülle test bench show strong correlation with simulation results. The study also introduces a methodology to compute the maximum admissible vehicle mass—determined to be 13.5 kg for a 180 W motor operating at 1900 rpm—based on acceleration and slope constraints. Furthermore, a switching algorithm for the bidirectional converter ensures reliable four-quadrant operation. Overall, the proposed framework provides a scalable and effective approach for EV emulation, control design, and energy management validation.

## I. INTRODUCTION

The intensifying climate crisis is primarily fueled by greenhouse gas emissions from Internal Combustion Engine (ICE) vehicles, which represent a significant source of environmental pollution. In response, the transportation sector has increasingly focused on developing energy-efficient and low-emission alternatives, with electric vehicles (EVs) emerging as a key sustainable solution. The shift toward EV technology now represents a dominant trend among major automotive manufacturers. Current research and development efforts are centered on designing transportation systems that are not only energy-efficient and environmentally friendly but also safe and reliable. These initiatives aim to address both ecological and societal concerns and are closely aligned with the United Nations Sustainable Development Goals (SDGs) [1], positioning EVs as vital components in building climate-resilient infrastructure.

Modern EVs demonstrate significantly lower emissions compared to ICE-powered vehicles, driven by advances in powertrain components that optimize performance, energy efficiency, and sustainability. These innovations align with global policy frameworks such as:

- European Union's 2035 ICE ban [2]
- U.S. Inflation Reduction Act's EV tax [3]

Moreover, enhancing EV systems remains an active area of research, focusing on improving vehicle range, refining control systems, and developing more efficient energy management strategies. In recent years, the concept of digital twins has emerged as a transformative tool, enabling the creation of virtual replicas of physical systems to optimize design, monitoring, and performance [4], [5].

The aim of this study is to develop a digital twin for a reduced-scale DC-drive electric vehicle (EV) emulator. Two modeling approaches are explored: one based on conventional block diagram representation, and the other utilizing the Energetic Macroscopic Representation (EMR) framework. Section II provides a review of related work, while Section III describes the implementation of the scaled-down EV emulator. Section IV presents the control architecture derived from the EMR model, along with validation results from both simulation and experimental testing.

## II. RELATED WORK

Modern EV powertrains [6] typically comprise a battery, an inverter, and an AC motor (see Fig. 1), with mechanical elements such as a differential and drive wheels transmitting torque to the road surface. In contrast, earlier EV models—such as the Peugeot 106, and Citroën Saxo—utilized DC motors combined with chopper-based control. Recently, there has been renewed interest in DC-based architectures. For example, [7] introduced an innovative four-quadrant DC chopper designed for series-excited motors, offering a cost-effective solution for DC drive applications in EVs.

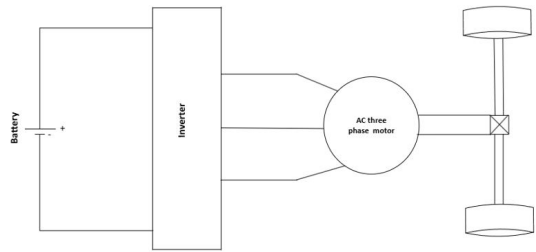

Fig. 1: Schematic representation of a modern EV

Recent research highlights the growing reliance on simulation-based approaches for EV design and optimization. In [8], the authors used MATLAB/Simulink-Simscape to convert a diesel-powered commercial vehicle into an electric version, incorporating a permanent magnet synchronous motor (PMSM). Likewise, study [9] investigated the impact of road slope on the dynamic behavior of a three-wheeled solar EV using MATLAB/Simscape modeling. Focusing on battery performance, study [10] employed Simcenter Amesim to simulate a Renault ZOE, validating the battery model with experimental data and analyzing overall vehicle performance.

*This work was not supported by any organization
[1]Icam School of Engineering, Lille Campus, 6 Rue Auber, B.P 10079, CEDEX, 59016 Lille, France `lamine.chalal@icam.fr`
[2] Laboratory of Innovative Technologies, University of Picardie Jules Verne, 80000 Amiens, France `rachid@u-picardie.fr`

In the context of advancing digital twin methodologies, study [11] introduced a simulation-only framework for EV propulsion systems, integrating a traction inverter with an embedded fast charger. While the model enables exploration of various operating modes to support powertrain electrification, it lacks experimental validation. Study [12], a MATLAB Simscape-based digital twin was developed for a solar race car, aimed at optimizing design and race strategies through cost-effective modeling; however, this work also did not include physical prototyping.

Study [13] stands out by combining physics-based models—such as photovoltaic panel and the motor— with a machine learning-based battery model to optimize energy management in solar vehicles. Crucially, this study validated the digital twin experimentally on a real solar car, effectively bridging the gap between simulation and real-world performance. Study [14] proposed a method for simulating EV braking behavior on a motor/dynamometer test bench, focusing on the allocation of regenerative and friction braking forces between the front and rear axles. Lastly, study [15] explored the development of Electronic Wedge Brake (EWB) technology using a hybrid experimental-simulation approach—including real-world tests, hardware-in-the-loop (HIL), and dynamometer-based evaluation—to reinforce the connection between simulation and practical validation.

## III. STUDIED REDUCED SCALE EV EMULATOR

To validate the feasibility of a digital twin for a scaled-down EV emulator and analyze its energy dynamics, this work implements a DC motor configuration driven by a four-quadrant (4Q) DC-DC converter. The system employs a SEDCM powered by the 4Q converter, interfaced with a battery emulator, as depicted in Fig. 2. Load torque is dynamically applied using a dynamometer integrated into the test bench, replicating real-world driving conditions. The SEDCM is mechanically coupled to a dynamometer, which emulates the resistive torque acting on the motor shaft (Fig. 3). This reference resisting torque is calculated using Newton's second law, incorporating external parameters such as vehicle weight, road gradient, and wind speed.

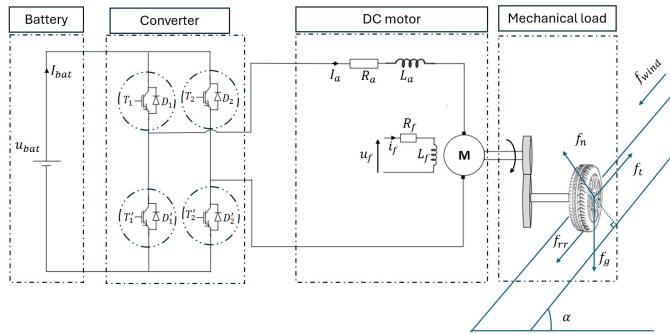

Fig. 2: SEDCM with 4Q DC-DC Converter and Battery Emulator

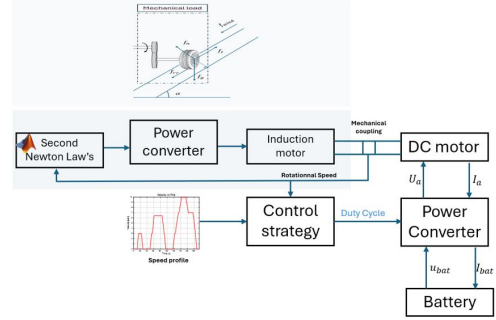

Fig. 3: Resistive Torque Emulation

### A. Mechanical Modelling

The vehicle's longitudinal dynamics are governed by the equilibrium between tractive effort and opposing forces. Resistive forces comprise aerodynamic drag ($F_{\text{drag}}$), rolling resistance ($F_{\text{roll}}$), and gradient resistance ($F_{\text{grad}}$) due to road inclination (slope angle $\alpha$). Applying Newton's Second Law in (1) yields the tractive force ($F_{Tr}$), produced by the electric powertrain. This force helps overcome the above-mentioned resistances.

$$F_{\text{Tractive}} = M \frac{dv}{dt} + \sum F_r \quad (1)$$

where:
- $F_{\text{Tr}}$: Longitudinal traction force at wheels [N]
- $\frac{dv}{dt}$: Vehicle acceleration [m/s$^2$]

The total resistance force combines three components as expressed by (2).

$$\sum F_r = \underbrace{Mg \sin \alpha}_{F_{\text{grad}}} + \underbrace{\text{sign}(v) Mg \mu_r \cos \alpha}_{F_{\text{roll}}} \\ + \underbrace{\text{sign}(v+w) \frac{1}{2} \rho C_d A_f (v+w)^2}_{F_{\text{drag}}} \quad (2)$$

where:
- $g = 9.81$ m/s$^2$: Gravitational acceleration
- $\alpha$: Road inclination angle [rad]
- $\mu_r$: Rolling resistance coefficient [-]
- $\rho = 1.2041$ kg/m$^3$: Air density at 20°C
- $A_f$: Frontal area [m$^2$]
- $v$: Vehicle speed [m/s]
- $w$: Headwind velocity [m/s]

To estimate the maximum achievable mass for a scaled-down EV powered by a 180 W DC motor operating at 1900 RPM, key system parameters were identified and incorporated into a structured computational workflow (see Fig. 4).

To assess the influence of acceleration and slope on this mass limit, a three-dimensional surface plot analysis was performed (Fig. 5).

Initialize parameters:
- High-speed shaft torque: $T_{r1} = 0.905 \text{ N}\cdot\text{m}$
- Reduction ratio: $K_R = 0.503$
- Wheel diameter: $D = 0.2$ m
- Max speed: $v = 10$ m/s
- Acceleration: $dv/dt = 1$ m/s$^2$

Compute transmission:
- Wheel torque: $T_{r2} = T_{r1}/K_R = 1.8 \text{ N}\cdot\text{m}$
- Max traction: $F_{Tr} = 2T_{r2}/D = 18$ N

Vehicle mass capacity (($\alpha = 0$, $w = 0$)):
$$M = \frac{F_{Tr} - \text{sign}(v+w)\cdot \frac{1}{2}\rho C_d A_f (v+w)^2}{a + g(\sin\alpha + \mu_r \cos\alpha)}$$
$$\approx \boxed{13.5 \text{ kg}}$$

Fig. 4: Flowchart for calculating vehicle mass capacity

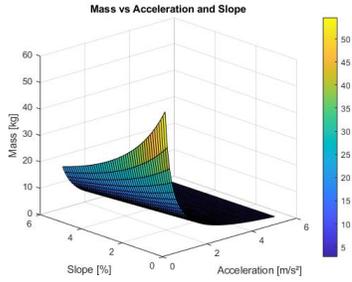

Fig. 5: Mass vs Acceleration and slope

The results reveal a clear inverse relationship: higher acceleration demands and steeper inclines significantly decrease the maximum allowable vehicle mass.

*B. SEDCM modelling*

This section details the mathematical modeling of a SEDCM. The motor's electrical configuration (Fig. 6) comprises two independent circuits: the field circuit and the armature circuit. Key electrical parameters are summarized in Table 1.

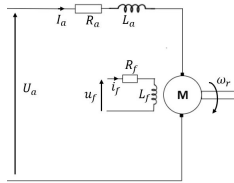

Fig. 6: Schematic of the SEDCM.

Applying Kirchhoff's voltage law to the field circuit yields:
$$u_f(t) = R_f i_f(t) + L_f \frac{d i_f(t)}{dt} \quad (3)$$

where:
- $u_f(t)$: Field circuit supply voltage (V)
- $R_f$: Field winding resistance ($\Omega$)
- $L_f$: Field inductance (H)
- $i_f(t)$: Field current (A)

For the armature circuit, the governing equation is:
$$U_a(t) = R_a I_a(t) + L_a \frac{d I_a(t)}{dt} + E_a(t) \quad (4)$$

where:
- $U_a(t)$: Armature supply voltage (V)
- $R_a$: Armature resistance ($\Omega$)
- $L_a$: Armature inductance (H)
- $I_a(t)$: Armature current (A)
- $E_a(t)$: Back-electromotive force (V)

For SEDCMs, the torque constant $k_m$ is proportional to the field current. This relationship can be expressed as:
$$k_m = K i_f \quad (5)$$

where $i_f$ represents the field current. In this configuration, the field current remains constant, making $k_m$ analogous to the back-electromotive force (back-EMF) constant observed in permanent magnet DC (PMDC) motors.

TABLE I: SEDCM Parameters

| Parameter | Value |
|---|---|
| Armature resistance ($R_a$) | $18.2\,\Omega$ |
| Armature inductance ($L_a$) | $380\,\text{mH}$ |
| Field resistance ($R_f$) | $2000\,\Omega$ |
| Field inductance ($L_f$) | $141\,\text{H}$ |
| Torque constant ($k_m$) | $0.91\,\text{N\,m/A}$ |
| Moment of inertia ($J$) | $0.0018\,\text{kg\,m}^2$ |

Speed regulation in SEDCMs employs two methods (Fig. 7):

1) **Armature Voltage Control** adjusts $U_a$ while keeping field current constant, enabling constant torque ($T_e = k_m I_a$) and linear mechanical power growth ($P_m = T_e \omega_r$) until base speed.
2) **Field Flux Weakening** reduces field current above base speed with nominal voltage $U_a$, extending speed range while torque decreases ($T_e \propto i_f$).

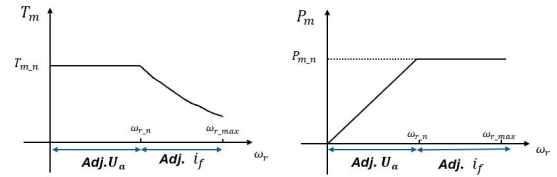

Fig. 7: Operating regions of a SEDCM.

The mechanical dynamics are governed by:
$$T_e(t) = k_m I_a(t) = T_0 + T_r(\Omega, t) + J\frac{d\Omega}{dt} \quad (6)$$

where:
- $T_r$: Load torque (N m)
- $T_0$: Friction and windage loss torque (N m)
- $J$: Combined moment of inertia (kg m$^2$)

*C. Power Converter Modeling*

To operate a DC motor at variable speeds, a variable voltage supply is required. This necessitates a power electronic converter that fulfills two key functions:
(i) Providing adjustable voltage to the motor input,

(ii) Enabling bidirectional operation (both current and voltage reversibility).

The proposed bidirectional DC/DC converter, which interfaces the battery emulator and an RL load (illustrated in Fig. 8), operates in four quadrants and incorporates four switches ($S_1$–$S_4$). Each switch comprises an IGBT (Insulated-Gate Bipolar Transistor) and an antiparallel diode, enabling:

- Voltage regulation capability,
- Seamless four-quadrant operation.

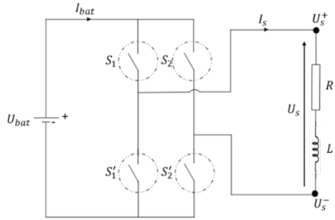

Fig. 8: Four-quadrant operation of the bidirectional converter

The primary control objective aims to regulate a positive average output voltage ($U_s$), particularly during the motoring mode in DC motor drives. As defined in Table II, switches $S_1$ and $S'_2$ are closed, enabling current flow from the battery emulator to the load. To ensure safe commutation, switch $S_2$ is enabled only after $S'_2$ is turned off. This technique facilitates the transfer of stored energy in the inductive load through switches $S_1$ and $S_2$. As depicted in Fig. 9, the voltage waveform across the load terminals and the current waveform are measured during operation with a positive average voltage $U_s$.

TABLE II: Switching states and current dynamics

| Closed Switches | $U_s$ | $I_s$ |
|---|---|---|
| $S_1 \& S'_2$ | $U_{\text{bat}}$ | Rising |
| $S_1 \& S_2$ | 0 | Falling |
| $S_1 \& S'_2$ | $U_{\text{bat}}$ | Rising |
| $S'_1 \& S'_2$ | 0 | Falling |

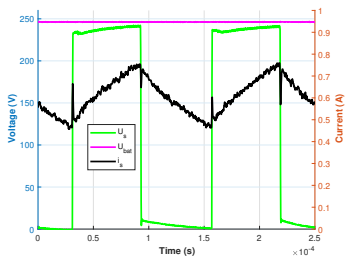

Fig. 9: Voltage and current waveforms for positive average output voltage

The average output voltage is calculated as:

$$\langle U_s \rangle = \frac{1}{T} \int_0^T U_s(t)\, dt$$
$$= \frac{1}{T}\left[\int_0^{\alpha T} U_{\text{bat}}\, dt + \int_{\alpha T}^T 0\, dt\right]$$
$$= \alpha U_{\text{bat}}$$

where $\alpha \in [0,\ 1]$, resulting in $\langle U_s \rangle \in [0,\ U_{\text{bat}}]$.

In accordance with the switching algorithm defined in Table III, switches $S_2$ and $S'_1$ are closed, enabling current flow from the battery emulator to the load and returning to the negative terminal of the input source. To ensure safe commutation, switch $S_1$ is enabled only after $S'_1$ is opened. Fig. 10 depicts the load terminal voltage and current waveforms under a negative average output voltage ($\langle U_s \rangle < 0$), particularly during reverse motoring mode.

TABLE III: Switching states and current behavior

| Closed Switches | $U_s$ | $I_s$ |
|---|---|---|
| $S_2 \& S'_1$ | $-U_{\text{bat}}$ | Falling |
| $S_2 \& S_1$ | 0 | Rising |
| $S_2 \& S'_1$ | $-U_{\text{bat}}$ | Rising |
| $S'_1 \& S'_2$ | 0 | Falling |

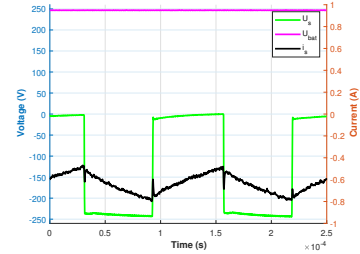

Fig. 10: Voltage and current waveforms for negative average output voltage

The average output voltage is expressed as:

$$\langle U_s \rangle = \frac{1}{T}\int_0^T U_s(t)\, dt$$
$$= \frac{1}{T}\left[\int_0^{\alpha T} -U_{\text{bat}}\, dt + \int_{\alpha T}^T 0\, dt\right]$$
$$= -\alpha U_{\text{bat}}$$

where $\alpha \in [\,0, 1]$. To extend this principle for bidirectional control, the duty cycle $\alpha'$ is generalized to both positive and negative values ($-1 \leq \alpha' \leq 1$). This allows the average output voltage to span both polarities, expressed as:

$$\langle U_s \rangle = \alpha' U_{\text{bat}}, \quad \alpha' \in [-1,\ 1] \quad (7)$$

This generalization establishes an average output voltage range of:

$$\langle U_s \rangle \in [-U_{\text{bat}},\ U_{\text{bat}}] \quad (8)$$

This averaged model relies on the assumption of idealized switching components, including diodes and IGBTs(Switching and conduction losses are neglected).

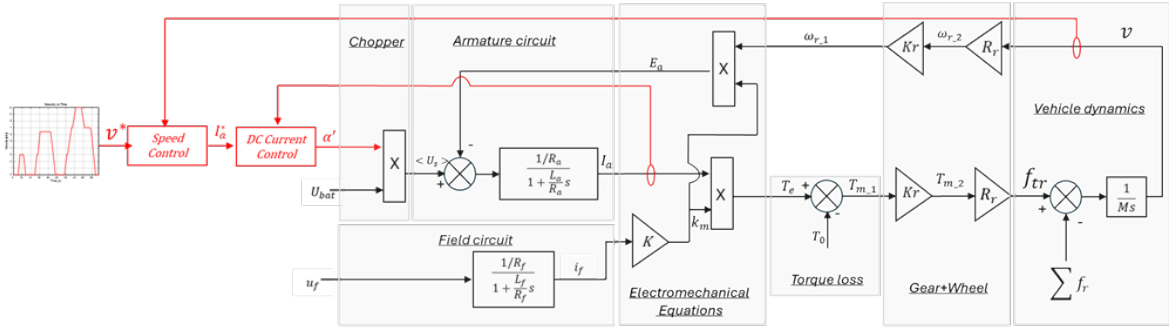

Fig. 11: Schematic of the Double-Loop Feedback Control System and SEDCM Block Diagram Model

## IV. CONTROL ARCHITECTURE AND VALIDATION

As illustrated in Fig. 11, the electric vehicle powertrain is modeled using a block diagram in the Laplace domain. The cascaded proportional-integral (PI) control structure for the separately excited DC motor (SEDCM), highlighted in red, consists of an outer speed control loop and an inner current control loop [16], [17]. This dual-loop configuration is designed to accurately follow the reference speed trajectory while mitigating the effects of external disturbances, such as variations in load torque. Although block diagrams are commonly used for representing linear systems, they present significant limitations when it comes to modeling the complex interactions and nonlinear behaviors characteristic of real-world systems [18].

To overcome these limitations, graphical modeling approaches such as Bond Graphs [19] and EMR [20] have been proposed for multiphysics system modeling. While Bond Graphs offer powerful tools for representing multidomain systems, their complexity can hinder control system design. In contrast, EMR provides a structured, control-oriented methodology that defines a functional representation of energy systems. It explicitly models energy exchanges between components based on principles of interaction and causality [20].

For the scaled-down vehicle studied, the EMR diagram—shown in Fig. 12— follows the direction of energy flow through the system:

- **Battery**: Modeled as an electrical source (green oval pictogram), characterized by variables $U_{bat}$ and $I_{bat}$.
- **4Q Chopper**: Represented by an orange square. Inputs: $U_{bat}$, $I_a$; Outputs: $U_{arm}$, $I_{bat}$. The variable of control is the duty cycle $\alpha'$.
- **Armature Circuit**: The circuit comprising armature resistance $R_a$ and inductance $L_a$ is depicted as crossed orange rectangle (accumulation element).
- **DC Motor**: functions as an electromechanical conversion element (orange circle). Inputs: $U_{arm}$, $\Omega_{1r}$; Outputs: $I_a$, torque $T_e$.
- **Gear and Wheel**: Mono-physical conversions (orange square), linking rotational speed $\Omega_{1r}$ to vehicle velocity $v_{ev}$.
- **Chassis**: Accumulation element (crossed orange rectangle) representing vehicle mass $m$, with velocity $v_{ev}$ as the state variable.
- **External Forces**: Resistive force $F_{res}$ (e.g., aerodynamics, slope) opposing motion.

The inversion-based control strategy is derived from the EMR using systematic inversion rules, as illustrated in the lower part of Fig. 12 (blue pictograms).

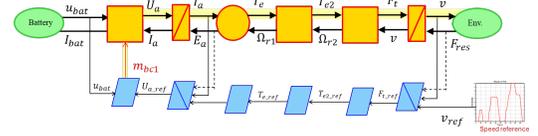

Fig. 12: EMR and Inversion-Based Control of the EV emulator

The inversion of the accumulation element representing the armature circuit requires closed-loop regulation. To track the reference current $I_{a\text{-ref}}$, the controller calculates the armature voltage $U_{a\text{-ref}}$ using:

- The real-time measured armature current $I_{a\text{-mea}}$,
- The estimated back-electromotive force $E_{a\text{-esti}}$.

To test and validate the current control loop, a current reference is applied. The experimental and simulation results demonstrate the validity of the implemented PI controller (see Fig. 13).

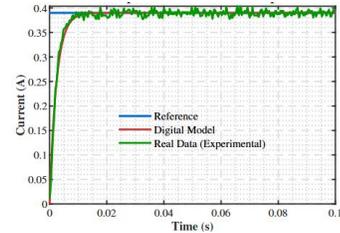

Fig. 13: Simulated vs Experimental Current Profiles

Similarly, the inversion of the accumulation element representing chassis dynamics requires closed-loop regulation. To track the reference vehicle velocity $v_{ev\text{-ref}}$, the controller calculates the traction force $F_{trac\text{-ref}}$ using:

- The real-time measured velocity $v_{ev\text{-mea}}$,
- The estimated resistive force $F_{res\text{-mea}}$, which accounts for slope angle, aerodynamic drag, and rolling resistance.

The remaining elements require direct inversion. The developed EMR model (Fig. 12) was implemented in MATLAB/Simulink. Experimental validation was performed using the DSP based Lucas-Nülle test bench to replicate real operating conditions of a reduced-scale EV (Fig. 14).

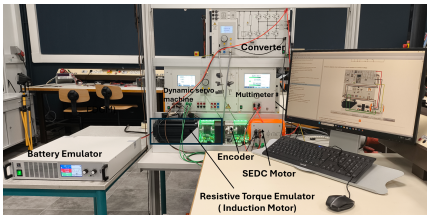

Fig. 14: Lucas-Nülle GmbH experimental setup

A comparative analysis of experimental and simulated speed profiles is presented in Fig. 15 demonstrating good correlation between measured and simulated results. A magnified view over the interval 60-80 s is included to highlight the correspondence during transient operation.

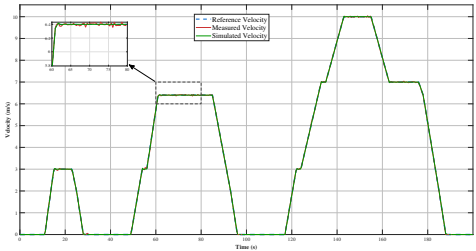

Fig. 15: Velocity tracking performance

## V. CONCLUSION

This study presents the development of a digital twin for a scaled-down EV emulator, incorporating detailed models of key subsystems—including a battery emulator, four-quadrant DC-DC converter, separately excited DC motor (SEDCM), and mechanical load—within the Energetic Macroscopic Representation (EMR) framework to systematically derive the control structure. A constraint-based methodology was introduced to determine the maximum permissible vehicle mass under realistic operational constraints (such as slope gradients, acceleration demands, and aerodynamic forces), thereby quantifying the impact of dynamic conditions on EV powertrain design. Experimental validation on a Lucas-Nülle test bench demonstrated strong correlation between simulation and measured data. This instrumented platform and its digital twin enable cost-effective testing of energy management strategies, while the modularity of the EMR approach supports hybridization studies (e.g., supercapacitor integration) for flexible architectural exploration. Future work will expand the framework to include AC motor inverter systems and a nonlinear battery model with state-of-charge (SOC) dynamics, aiming to optimize EV autonomy across various driving profiles (urban, highway) and traffic conditions. Validation on a full-scale platform is also planned to evaluate the scalability and real-world applicability of the proposed methodology.